\documentclass[aps,prd,twocolumn,nofootinbib,superscriptaddress,showpacs,amsmath,amssymb]{revtex4}
\usepackage{graphicx,epsf}
\usepackage[]{latexsym}
\newcommand{\ve}{\varepsilon}
\newcommand{\be}{\begin{eqnarray}}
\newcommand{\ee}{\end{eqnarray}}
\newcommand{\bea}{\begin{eqnarray}}
\newcommand{\eea}{\end{eqnarray}}

\def\comment#1{}

\newcommand{\lp}{\ell_{\rm p}}
\newcommand{\mpl}{m_{\rm p}}

\usepackage{color}

\definecolor{darkred}{rgb}{.8,0,0}

\definecolor{darkblue}{rgb}{0,0,.7}

\definecolor{darkgreen}{rgb}{0,.7,0}

\begin{document}

%
%
\title{Lorentz violation and generalized uncertainty principle}

\author{Gaetano Lambiase}
\email{lambiase@sa.infn.it}
\affiliation{Dipartimento di Fisica "E.R. Caianiello", Universita' di Salerno, I-84084 Fisciano (Sa), Italy \&\\
INFN - Gruppo Collegato di Salerno, Italy}
\author{Fabio Scardigli\footnote{corresponding author}}
\email{fabio@phys.ntu.edu.tw}
\affiliation{Dipartimento di Matematica, Politecnico di Milano, Piazza Leonardo da Vinci 32, 20133 Milano, Italy}
\affiliation{Department of Applied Mathematics, University of Waterloo, Ontario N2L 3G1, Canada}
\begin{abstract}
\par\noindent
Investigations on possible violation of Lorentz invariance have been widely pursued in the last decades, both from theoretical and experimental sides. A comprehensive framework to formulate the problem is the standard model extension (SME) proposed by A.Kostelecky, where violation of Lorentz invariance is encoded into specific coefficients. Here we present a procedure to link the deformation parameter $\beta$ of the generalized uncertainty principle (GUP) to the SME coefficients of the gravity sector. The idea is to compute the Hawking temperature of a black hole in two different ways. The first way involves the deformation parameter $\beta$, and therefore we get a deformed Hawking temperature containing the parameter $\beta$. The second way involves a deformed Schwarzschild metric containing the Lorentz violating terms $\bar{s}^{\mu\nu}$ of the gravity sector of the SME. The comparison between the two different techniques yields a relation between $\beta$ and $\bar{s}^{\mu\nu}$. In this way bounds on $\beta$ transferred from $\bar{s}^{\mu\nu}$ are improved by many orders of magnitude when compared with those derived in other gravitational frameworks. Also the opposite possibility of bounds transferred from $\beta$ to $\bar{s}^{\mu\nu}$ is briefly discussed.
\end{abstract}
%

%
\maketitle
\section{Introduction}

Possible breakdowns of the fundamental symmetries in Physics have received a more and more growing
interest and have been studied in different areas (see for example \cite{kostelecky,ellis,loop,NCG,bertolami,mattingly}). The most general setting in which they have been investigated is the Standard Model Extension (SME) \cite{kostelecky}. The violation of the fundamental symmetries, i.e. Lorentz's and CPT symmetries, follows from the observation that the vacuum solution of the theory could spontaneously violate them, even though they are preserved by the underlying theory. Modern tests for Lorentz and CPT invariance
breakdown have been discussed in \cite{mattinglytests}. More recently, the SME has been extended to incorporate the gravitational interaction \cite{kostelecky,bluhm,kosteleckyG}. The latter results  foresee that the effective action is given by
 \[
S=S_{HE}+S_m+S_{LV}\,,
 \]
where $S_{HE}=(16\pi G)^{-1}\int d^4x \sqrt{-g}(R-2\Lambda)$ is the standard Hilbert-Einstein action of General
Relativity ($\Lambda$ is the cosmological constant), $S_m$ the general matter action (which also includes Lorentz violating matter gravity coupling), while Lorentz violating gravitational couplings are included in $S_{LV}$ \cite{PPN}
\begin{equation}
\label{LSaction}
S_{LV}=\frac{1}{16\pi G}\int d^4e\left(-uR+s^{\mu\nu}R^T_{\mu\nu}+t^{\kappa\lambda\mu\nu}
C_{\kappa\lambda\mu\nu}\right)\,,
\end{equation}
where $R^T$ is the trace-free Ricci  tensor and $C_{\kappa\lambda\mu\nu}$ is the Weyl conformal tensor.
The coefficients $u$, $s^{\mu\nu}$ and $t^{\kappa\lambda\mu\nu}$ are real and dimensionless. The coefficients $s^{\mu\nu}$ and $t^{\kappa\lambda\mu\nu}$ fulfill the Ricci and Riemann properties, respectively, and are traceless:
 \[
 s^{\mu}_{{\phantom \mu} \mu}=0\,, \quad
 t^{\kappa\lambda}_{\phantom{\kappa\lambda}\kappa\lambda}=0\,, \quad
 t^{\kappa}_{\phantom{\kappa}\mu\kappa\lambda}=0\,.
 \]
We restrict to the case $u=0$ and $t^{\kappa\lambda\mu\nu}=0$, therefore only the coefficients $s^{\mu\nu}$ control the Lorentz violation degrees of freedom. By varying the action $S$ with respect to the background metric yields
\cite{PPN}
\begin{equation}\label{fieldequations}
  G^{\mu\nu}-(T^{Rs})^{\mu\nu}=8\pi G\, T_g^{\mu\nu}\,,
\end{equation}
where $G^{\mu\nu}=R^{\mu\nu}-(R/2)g^{\mu\nu}$ is the standard
Einstein tensor, and
\be\label{TRs}
(T^{Rs})^{\mu\nu} &=& \frac{1}{2}g^{\mu\nu}(R_{\alpha\beta}-
  \nabla_\alpha \nabla_\beta) s^{\alpha\beta} \\
  & + &   \frac{1}{2}\left(\nabla_\alpha \nabla^\mu s^{\alpha\nu}+\nabla_\alpha \nabla^\nu s^{\alpha\mu} -\nabla_\alpha \nabla^\alpha s^{\mu\nu}\right)\,. \nonumber
\ee
The PPN approximation of (\ref{LSaction}) has been studied in \cite{PPN}.
There a method is developed to extract the modified Einstein field equations in the limit of small metric fluctuations above the Minkowski vacuum, while allowing for the dynamics of 20 independent coefficients for Lorentz violation. The linearized effective equations are solved to obtain the post-Newtonian metric. Then the equations of motion for a perfect fluid in this metric are obtained, and applied to a many body gravitating system. Finally, tidal forces are disregarded, and the point particle limit of these equations is considered. This procedure yields a two point-particles Lagrangian, which gives the effective equations of motion for the two bodies system, in the coordinate acceleration. In the hypothesis $M \gg m$ and considering the heaviest body $M$ at rest (with respect to the test particle with mass $m$), this effective two bodies Lagrangian reads
\be
L &=& \frac{1}{2}m v^2 + \frac{G m M}{r}\left(1 + \frac{3}{2}\bar{s}^{00} +
\frac{1}{2} \bar{s}^{jk}\frac{x_j x_k}{r^2}\right) \nonumber \\
&-& \frac{G m M}{r}\left(3 \bar{s}^{0j}v_j + \bar{s}^{0j}\frac{x_j}{r}v_k\frac{x_k}{r}\right)
\label{KL}
\ee
where we understood summation over indexes $j, k$, and
where $v^2=v_1^2 + v_2^2 + v_3^2$, $r^2 = x_1^2 + x_2^2 + x_3^2$, $v_k = \dot{x_k}$, and the derivative is taken in respect to the coordinate time. The $\bar{s}^{\mu\nu}$ are the Lorentz violation coefficients.
We work in units with $c=1$, and we consider particles moving slowly in respect to the speed of light, in a stationary and weak gravitational field. Therefore $v \ll 1$, so at the level of approximation we need, we can neglect the terms depending on the velocity $v$. The effective potential therefore reads
\be
V(r)=\frac{U(r)}{m}= -\frac{GM}{r}\left(1 + \frac{3}{2}\bar{s}^{00} +
\frac{1}{2} \bar{s}^{jk}\frac{x_j x_k}{r^2}\right)
\label{LV1}
\ee
As indicated in several points of Ref.~\cite{PPN} (see Eqs. 62, 100, 126, 137, 175) the scalar factor
$(1 + 3\bar{s}^{00}/2)$ merely acts as a rescaling of the gravitational constant, and hence is unobservable in the present context. We can in fact rewrite $V(r)$ as
\be
V(r) = -\frac{G_{\rm eff}M}{r}\left(1 +  \bar{s}^{jk}_{\rm eff}\frac{x_j x_k}{r^2}\right)
\label{LV2}
\ee
where
\be
G_{\rm eff}=G\left(1 + \frac{3}{2}\bar{s}^{00}\right) \ ; \quad \bar{s}^{jk}_{\rm eff}=\frac{\bar{s}^{jk}}{2+3\bar{s}^{00}}
\ee
Notice that the term $\bar{s}^{jk}x_j x_k/r^2$ cannot be reabsorbed into $G_{\rm eff}$ since it is an anisoptropic term, which depends upon the directions $x_j x_k/r^2$.
Given the unobservability of the factors containing $\bar{s}^{00}$, in the following we shall simply rename $G_{\rm eff}\equiv G$ and $\bar{s}^{jk}_{\rm eff}\equiv \bar{s}^{jk}$. The parameters $s^{\mu\nu}$ have been constrained in different frameworks, see Table \ref{Tab01} \cite{kostRMP2013}
   \begin{center}
\begin{table}[ht]
\caption{Upper bounds on the ${\bar s}^{\mu\nu}$ derived in different physical frameworks (see Ref. \cite{kostRMP2013} for
a complete list).}
\begin{tabular}{c|c|c}
  \hline\hline
  ${\bar s}^{\mu\nu} <$\quad  & \qquad Physical framework \qquad  & \quad Refs.  \\ \hline
 $10^{-15}$ & Torsion pendulum & \cite{PPN} \\
  & & \\
 $10^{-14}$ & Cosmic rays  & \cite{[21]} \\
   & & \\
 $10^{-12}$ & Lunar laser ranging &  \cite{[241]} \\
  & & \\
 $10^{-11}$ & Binary pulsars & \cite{[245]} \\
 & & \\
 $10^{-9}$ & Atom interferometry & \cite{[250]}  \\
           & Perihelion precession & \cite{PPN}     \\
           & (Solar system data) &                     \\ \hline
 \end{tabular}
\label{Tab01}
\end{table}
\end{center}

The aim of this paper is to relate the coefficients ${\bar s}^{\mu\nu}$ to the  the deforming parameter $\beta$ of the generalized uncertainty principle (GUP) \cite{GUPearly}. In fact, one of the most studied forms of the deformation of Heisenberg uncertainty principle (HUP), usually called GUP, is
\be
\mbox{\hspace{-5mm}}\Delta x\, \Delta p &\geq&
\frac{\hbar}{2}\left(1  +  \beta\,\frac{4\,\lp^2}{\hbar^2}\,\Delta p\,^2\right) 
=
\frac{\hbar}{2}\left[ 1 + \beta \left(\frac{\Delta p}{\mpl} \right)^2\right]
\label{gup}
\ee
%
Here $l_p$ is the Planck length, $m_p$ the Planck mass, and we work in units where $2Gm_p=l_p$, $\hbar=2m_pl_p$, $c=k_B=1$. Typically, investigations mainly focus on understanding how gravity may affect the formulation of Heisenberg Uncertainty Principle (HUP). Given the pivotal role of gravitation in these arguments, it is not surprising that the most relevant modifications to the HUP have been proposed in string theory, loop quantum gravity, deformed special relativity, and studies of black hole physics~\cite{VenezGrossMende,MM,kempf,FS,Adler2,SC}. In principle, the dimensionless parameter $\beta$
is not fixed by the theory, although it is generally assumed to be of order one (this happens, in particular, in some models of string theory Ref.~\cite{VenezGrossMende}).

In our approach we use the GUP to compute the Hawking temperature of a given black hole, which however can be also computed as well via the effective potential $V(r)$, and hence related to ${\bar s}^{\mu\nu}$. We shall find $\beta\simeq (M/m_p)^2|{\bar s}^{jk}|$, where $M$ is the mass of the gravitational source. By making use of the most stringent bound on the parameter ${\bar s}^{jk}$ (see Table \ref{Tab01}), we will derive an upper bound on the deformation parameter $\beta$.

\section{GUP-deformed Hawking temperature}

As is well known from the argument of the Heisenberg microscope~\cite{Heisenberg},
the size $\delta x$ of the smallest detail of an object, theoretically detectable
with a beam of photons of energy $E$, is roughly given by
\be
\delta x
\simeq
\frac{\hbar}{2\, E}
\label{HS}
\ee
since ever larger energies are required to explore ever smaller details.
From the uncertainty relation~\eqref{gup}, we see that the GUP version of the standard
Heisenberg formula~\eqref{HS} is
\be
\delta x
\simeq
\frac{\hbar}{2\, E}
+ 2\,\beta\,\lp^2\, \frac{E}{\hbar}
\label{He}
\ee
which relates the (average) wavelength of a photon to its energy $E$
\footnote{Here, the standard dispersion relation $E=p\,c$ is assumed.}.
To compute the thermal GUP corrections to the Hawking spectrum, we follow the arguments of Refs.~\cite{FS9506,ACSantiago,CavagliaD,CDM03,Susskind,nouicer,Glimpses,SLV}.
We can derive from (\ref{He}) a relation between the mass and the temperature of a Schwarzschild black hole.
Consider an ensemble of unpolarized photons of Hawking radiation
just outside the event horizon of a Schwarzschild black hole. From a geometrical point of view,
it is easy to see that the position uncertainty of such photons is of
the order of the Schwarzschild radius $R_S=2GM$ of the black hole.
Hence, the photon
positional uncertainty is
$\delta x  \simeq  2\mu R_S$.
%
%
The proportionality constant  $\mu$ is of order unity, as we will see.
%
%
According to the equipartition principle, the average energy
$E$ of unpolarized photons of the Hawking radiation is simply related with their temperature $T$ by
$E \ = \ T$.
Formula (\ref{He}) then becomes
\be
4\mu GM \ \simeq \ \frac{\hbar }{2 T} \ + \ 2\beta G T \,.
\label{37}
\ee
Finally, we have
\be
M \ = \ \frac{\hbar }{8\pi G  T} \ + \ \beta\, \frac{ T}{2\pi }
\ee
where we fixed $\mu = \pi$ by requiring that formula (\ref{37}) predicts the standard semiclassical
Hawking temperature, when the semiclassical limit $\beta \to 0$ is considered.

This is the black hole mass-temperature relation predicted by the GUP for a Schwarzschild black hole.
Of course this relation can be easily inverted, to get $T=T(M)$. Since however the term in $\beta\, T$ is small, especially for solar mass black holes, it is more useful to invert and expand in powers of $\beta$.
We arrive to the expression

\be
T = \frac{\hbar}{8\pi GM} \left(1 + \frac{\beta\,\mpl^2}{4\pi^2\,M^2} + \dots \right)~.
\label{Tg}
\ee
and it is evident that to zero order in $\beta$, we recover the usual well known Hawking formula.
We stress that we are assuming that the correction induced by the GUP has
a thermal character, and, therefore, it can be cast in the form of a shift of the Hawking temperature. Of course, there
are also different approaches, where the corrections do not respect the exact thermality of the spectrum, and thus need
not be reducible to a simple shift of the temperature (an example is the corpuscular
model of a black hole of Ref.~\cite{dvali}).

\section{Metric mimicking a potential corrected with Lorentz violating terms}

Now we  consider the effective potential produced by a metric of the very general class
\be
ds^2 = F(r)dt^2 - g_{ik}(x_1,x_2,x_3) dx^i dx^k
\label{ik}
\ee
where $r=|\mathbf x|=(x_1^2 + x_2^2 + x_3^2)^{1/2}$, and $x_1,x_2,x_3$ are the standard Cartesian coordinates.
Particular cases of the metric (\ref{ik})  is  the Schwarzschild metric, in the standard form
\be
ds^2 = \left(1-\frac{2GM}{r}\right)dt^2 - \left(1-\frac{2GM}{r}\right)^{-1}dr^2 - r^2d\Omega^2 \nonumber
\ee
as well as in harmonic coordinates
\be
ds^2 &=& \left(\frac{R-GM}{R+GM}\right)dt^2 - \left(\frac{R+GM}{R-GM}\right)dR^2 \nonumber \\
&-& (R+GM)^2 d\Omega^2 \,.\nonumber
\ee
with  $R=r-GM$.
\par\noindent
It can be easily seen \footnote{More details can be found in Ref.  \cite{Weinberg72}.}
that any general metric of the form
\be
ds^2 = F(r)dt^2 - F(r)^{-1}dr^2 - C(r)d\Omega^2
\label{gm}
\ee
can be put in the form (\ref{ik}).
In fact, Eq. (\ref{gm}) is equivalent to
\begin{eqnarray}
ds^2 &=& F(r)dt^2 - \left(F(r)^{-1} - \frac{C(r)}{r^2}\right)\frac{1}{r^2}({\bf x} \cdot d{\bf x})^2 \nonumber \\
 &-& \frac{C(r)}{r^2}d{\bf x}^2\,. \nonumber
\end{eqnarray}
Once the metric is in the form (\ref{ik}), in Cartesian coordinates, then, with well known procedures  \cite{Weinberg72},
it is easy to show that the effective Newtonian potential \footnote{The effective Newtonian potential is produced
by the metric given  in (\ref{ik}) for  a point particle which  moves slowly, in a stationary  and weak gravitational
field, i.e., quasi-Minkowskian far from the source, $r \to \infty$.} is of the form
\be
V(r) \ \simeq \ \frac{1}{2} \ (F(r)-1)
\ee
or,  equivalently,
\be
F(r) \ \simeq \ 1 \ + \ 2 \ V(r)~.
\ee
Therefore, the metric able to mimic the corrected
Newtonian potential (\ref{LV2}), containing Lorentz violating terms, will be
\be
F(r) = 1 -\frac{2GM}{r}\left(1 + \bar{s}^{jk}f_{jk}(\theta, \phi)\right).
\label{LVM}
\ee
where we have introduced standard spherical coordinates ${\bf x}=r (\sin\theta\cos\phi, \sin\theta\sin\phi, \cos\theta)$ and
$x_j x_k / r^2 = f_{jk}(\theta, \phi)$.
Notice that the formal angular dependence on $(\theta, \phi)$ displayed by (\ref{LVM}), in our context does not really matter. In fact we know that $|\bar{s}^{\mu\nu}|\lesssim 10^{-10}$, $|f_{jk}(\theta, \phi)|\leq 1$ for any $(\theta, \phi)$, and we are just interested in transferring bounds from $\bar{s}^{jk}$ to $\beta$. Hence, angular dependence does not influence our considerations.

\section{Temperature from a deformed Schwarzschild metric}

Let us define, for fixed $\theta$ and $\phi$,
\be
\epsilon(r) = -\frac{2GM}{r}\,\bar{s}^{jk}f_{jk}(\theta, \phi).
\label{phi}
\ee
Therefore, $F(r)$ will now be of the form
\be
F(r) = 1 - \frac{2GM}{r} + \epsilon(r).
\label{DM}
\ee
Since $|f_{jk}(\theta, \phi)| \leq 1$, and we are supposing $|\bar{s}^{jk}| \ll 1$, then
it is clear that $|\epsilon(r)| \ll 2GM/r$ for any $r\geq 2GM$.

We can legitimately wonder what kind of deformed Hawking temperature
can be inferred from a deformed Schwarzschild metric as in (\ref{DM}).
The deformation (\ref{DM}) makes sense when $|\epsilon(r)| \ll GM/r$.
For computational reasons, we can introduce a regulatory small parameter $\ve$ so that we can write
$\epsilon(r) \equiv \ve\phi(r)$. At the end of the calculation, $\ve$ can can be sent to unity.
Of course, we will look for the lowest order correction in the dimensionless parameter $\ve$.
The horizon's equation, i.e., $F(r)=0$, now reads
\be
r - 2 \,G\, M + \ve\, r\, \phi(r) = 0~.
\label{eqr}
\ee
Such equations  can be solved, in a first approximation in $\ve$. The solution reads (see Appendix A)
\be
r_H \ = \ a - \frac{\ve\, a\, \phi(a)}{1 \ + \ \ve\,[\phi(a) \ + \ a\,\phi'(a)]}
\label{solr}
\ee
where $a = 2 G M$.
Using Eqs.(\ref{phi}) and (\ref{DM}) for $F(r)$, and Eq.(\ref{solr}) for $r_H$, and expanding in $\ve$, one arrives to
the deformed Hawking temperature as (see Appendix B)
\be
T &=& \hbar \frac{F'(r_H)}{4\pi} = \frac{\hbar}{4\pi a} \left\{1 + \ve \left[2\phi(a) + a\phi'(a)\right] \right. \nonumber \\
&+& \left. \ve^2 \phi(a) \left[\phi(a) - 2a\phi'(a) - a^2 \phi''(a) \right] + \dots \right\}.
\label{teps}
\ee
It is noteworthy that the only function $\phi(r)$ that  annihilates the first-order in $\ve$  temperature
correction term is the solution of the differential equation $2\phi(r) + r\phi'(r) \ = \ 0$,
namely $\phi(r) = A / r^2$, where $A$ is an arbitrary constant.
In particular, for the function $\phi(r) = G^2 M^2 / r^2$, the coefficient of $\ve$ in (\ref{teps}) is zero,
and the coefficient of $\ve^2$ is $-1/16$.

\section{Relation between $\beta$ and $\bar{s}^{jk}$}


We are now in the position to compute the temperature generated by the metric (\ref{DM}), by simply
employing  (\ref{teps}). Therefore,  the metric-deformed Hawking temperature  is of the form
\be
T &=&\frac{\hbar}{4\pi a} \left\{1 + \left[2\epsilon(a) + a\epsilon'(a)\right] + \dots \right\}
\label{tesp2}
\ee
while the GUP-deformed Hawking temperature reads
\be
T = \frac{\hbar}{8\pi GM} \left(1 + \frac{\beta\,\mpl^2}{4\pi^2\,M^2} + \dots \right)~.
\ee
By comparing the two respective first-order correction terms in the two previous expansions, we obtain
\be
\beta \ = \  \frac{4\pi^2 M^2}{\mpl^2}\, \left[2\epsilon(a) + a\epsilon'(a)\right]~.
\label{beta1}
\ee
Using now the expression (\ref{phi}) for $\epsilon(r)$, we get
\be
\beta = -\frac{4\pi^2 M^2}{\mpl^2}\,\bar{s}^{jk}f_{jk}(\theta, \phi)
\label{beta}
\ee
for fixed\footnote{We note that the relation (\ref{beta}) can be derived also in a different way. The metric (\ref{LVM}) can be written as
\be
F(r) = 1-\frac{2GM}{r} \cdot A \nonumber
\label{MA}
\ee
where $A=1+\xi$, with $\xi=\bar{s}^{jk}f_{jk}(\theta, \phi)$. The horizon $F(r)=0$ is now $r_H = 2GMA$, and therefore the temperature is $T = \hbar \frac{F'(r_H)}{4\pi}  \simeq \frac{\hbar}{8 \pi GM}(1-\xi)$. Finally, on comparing with (\ref{Tg}) we get $\beta = -\frac{4\pi^2 M^2}{\mpl^2}\,\xi$, which coincides with (\ref{beta}).} $\theta$ and $\phi$.

Again, we can comment that angular dependence shown by Eq.(\ref{beta}) is not particularly puzzling in this context, since we are here interested in linking the magnitudes of $\bar{s}^{jk}$ and $\beta$, therefore it suffices to notice that $|f_{jk}(\theta,\phi)|\leq 1$ for any $(\theta,\phi)$.
Furthermore, the fact that $\beta$ in some cases could result \emph{negative} for positive values of $\bar{s}^{jk}$ (in general, the quantities $\bar{s}^{jk}$ can be positive or negative), shouldn't actually be a worry, since negative $\beta$ can be interpreted as a signal of a lattice structure of the space-time at the Planck scale (see Refs.~\cite{JKS,SC2}).
We can get rid of angular dependence by averaging over $(\theta, \phi)$, in fact
$\langle x^j x^k\rangle =r^2 \delta^{jk}/3$, and therefore setting $f_{jk}=\frac{1}{3}\delta_{jk}$, one gets (see (\ref{beta}))
\be
 \beta= \frac{4\pi^2}{3}\left(\frac{M}{\mpl}\right)^2\delta\,, \quad \delta \equiv |\bar{s}^{11}+\bar{s}^{22}+\bar{s}^{33}|=|\bar{s}^{ii}|\,.
\label{betaftrace}
\ee

\section{Comparison with experimental data.}

\subsection{Bounds from $\bar{s}^{jk}$ to $\beta$}

Before proceeding further, we have to clarify what kind of mass $M$ we have to insert in formula (\ref{betaftrace}) in order to compute the relevant bounds on $\beta$. 
Clearly, $M$ is the same mass which appears in the deformed metric (\ref{LVM}). Therefore $M$ will be chosen according to the specific gravitational experiment used to produce specific bounds (of gravitational origin) on $\bar{s}^{jk}$.
Examining the experimental situations described in Refs.~\cite{PPN,[241],[245]} (see there in particular Table VI of \cite{PPN}), and the relevant gravitational bounds on $\bar{s}^{jk}$, we can list the following cases.

\textbf{Lunar ranging.} It is well known that lunar laser ranging is among the most sensitive tests of gravitational physics within the solar system. In this situation $M$ is the mass of the Earth, and the distance of the Moon is probed with lasers with a precision at the centimeter level. According to Ref.~\cite{PPN} and its Table VI, the best attainable sensitivity is of order $|\bar{s}^{jk}| \lesssim 10^{-11}$. According to Ref.\cite{[241]}, we have $|\bar{s}^{jk}| \lesssim 10^{-12}$. In any case here we have $M=M_{Earth}\simeq 2.74 \times 10^{32}\mpl$.

\textbf{Binary pulsars.} In the case of binary pulsars we have bounds on the quantities $\bar{s}_e$, $\bar{s}_{\omega}$. In particular $\bar{s}_{\omega} \leq 10^{-11}$. According to Ref.~\cite{PPN}, both $\bar{s}_e$, $\bar{s}_{\omega}$ are linear combinations of the quantities $\bar{s}_{PP}$, $\bar{s}_{kP}$, $\bar{s}_{QQ}$, etc. (see Eq.184 of ~\cite{PPN}), which in turns are linear combinations of the quantities $\bar{s}^{jk}$ (see Eq.185 of ~\cite{PPN}). The coefficients entering these linear combinations are all of order 1, therefore the bounds stated for $\bar{s}_{\omega}$ can be safely transferred to $\bar{s}^{jk}$. So finally we can affirm that from binary pulsar data we get the bounds $|\bar{s}^{jk}| \leq 10^{-11}$, confirmed by Ref.\cite{[245]}. Here the relevant mass is the total mass of the binary pulsar system (on this, see also Refs.~\cite{SC2,TW,TW89}), namely $M=m_1+m_2$. Considering the very well known system PRS B 1913+16, we have $M=m_1+m_2=2.828\times M_{\bigodot}=2.55\times 10^{38} \mpl$.

\textbf{Perihelion precession.} For the perihelion precession in the solar system we consider of course in particular the data from Mercury. Again we see from Ref.~\cite{PPN}, Eq.191, that we have a bound $\bar{s}_{Mer} \leq 10^{-9}$, and $\bar{s}_{Mer}$ turns out to be a linear combination of $\bar{s}^{jk}$ with coefficients $O(1)$. So we can use that bound also for $\bar{s}^{jk}$. Obviously, the mass here involved is the mass of the Sun $M=M_{\bigodot}=0.9\times10^{38}\mpl$.

\textbf{Torsion pendulum.} We conclude this inspection with the most stringent case, as for the bounds on $\bar{s}^{jk}$. In a laboratory experiment, on the Earth surface, a torsion pendulum has been considered in Ref.~\cite{PPN}. The bounds in principle attainable with this device on the coefficients $\bar{s}^{JK}$ are of order $10^{-15}$ (see again table VI of
Ref.~\cite{PPN}). The quantities $\bar{s}^{JK}$ are here again just linear combinations of the quantities $\bar{s}^{jk}$, with coefficients of $O(1)$. So we can infer the bounds $|\bar{s}^{jk}|\lesssim 10^{-15}$. Obviously the mass to be considered here is again the Earth mass (see Eqs.119, 121, 122 of Ref.~\cite{PPN}), namely $M=M_{Earth}\simeq 2.74 \times 10^{32}\mpl$.

The above analysis suggests that the relevant masses to be inserted in relation (\ref{betaftrace}), in order to get bounds on $\beta$, are essentially the Earth mass, $M_{Earth}\simeq 2.74 \times 10^{32}\mpl$, or the Solar mass, $M_{\bigodot}=0.9\times10^{38}\mpl$.

As for the constraints on the parameter $\beta$, in recent years there has been a wide and lively research on this topic (see e.g. Ref.\cite{feng17,brukner,martin}), summarized in Tables \ref{Tab2} and \ref{Tab3}.
To be consistent with the logic of the argument presented here, we focus only on the bounds of $\beta$ of gravitational origin,  reported in Table \ref{Tab2}.
If we use now, in relation (\ref{betaftrace}), the best bound on ${\bar s}^{jk}$, namely the one from Torsion Pendulum experiments, $|{\bar s}^{jk}|<10^{-15}$, and consequently the Earth mass for $M=M_{Earth}$, then we get for the deformation parameter of GUP
\begin{equation}\label{betaboundSME}
\beta < 10^{51}\,.
 \end{equation}
As it clear from Table \ref{Tab2}, this bound improves by many orders of magnitude almost all the bounds on $\beta$ inferred in different gravitational experiments. In particular, the procedure above outlined does not involve a violation of the equivalence principle. Finally, the bound (\ref{betaboundSME}) is quite close to the one derived from Landau levels measurements, i.e. a non-gravitational bound.
\begin{center}
\begin{table}[ht]
\caption{Upper bounds on the GUP parameter $\beta$ inferred in gravitational measurements/experiments.}
\begin{tabular}{c|c|c}
  \hline\hline
  $\beta <$ & Physical framework  & Refs.  \\ \hline
 $10^{21}$ & Violation of equivalence principle & \cite{ghosh} (2014)\\
           & (on Earth) & \\
           &  Law of reciprocal action  &  \\ \hline
 $10^{60}$ & GW 150914 & \cite{feng17} \\ \hline
 $10^{69}$ & Perihelion precession &  \cite{SC2} (2015) \\
            & (Solar system data)  &     \\  \hline
 $10^{71}$ & Perihelion precession  & \cite{SC2} (2015) \\
            & (Pulsar PRS B 1913+16 data) &  \\ \hline
 $10^{78}$ & Modified mass-temperature relation & \cite{SC2} (2015) \\
           & Light deflection &     \\    \hline \hline
 \end{tabular}
\label{Tab2}
\end{table}
\end{center}
\begin{center}
\begin{table}[ht]
\caption{Upper bounds on the GUP parameter $\beta$ inferred in different non-gravitational measurements/experiments.}
\begin{tabular}{c|c|c}
  \hline\hline
  $\beta <$ & Physical framework  & Refs.  \\ \hline
 $10^{18}$ & Evolution of micro and nano  & \cite{Bonaldi} (2015)\\
             &  mechanical oscillators (masses $\sim \mpl$)  & \\ \hline
 $10^{20}$ & Lamb shift &  \cite{vagenas} (2011) \\ \hline
 $10^{21}$ & Scanning tunneling microscope & \cite{vagenas} (2008) \\ \hline
 $10^{33}$   & Gravitational bar detectors\footnote{This bound is derived without explicitly involving the gravitational interaction.} & \cite{martin} (2013) \\ \hline
 $10^{34}$  & Electroweak measurement & \cite{vagenas} (2011) \\ \hline
 $10^{34}$ & Charmonium levels & \cite{vagenas} (2011) \\
           & Energy difference in Hydrogen & \cite{Tkachuk} (2010)\\
           & levels $1S-2S$                  & \\ \hline
 $10^{39} $ & ${}^{87}$Rb cold-atom-recoil experiment & \cite{gao} (2016) \\ \hline
 $10^{46}$ & Landau levels & \cite{vagenas} (2011) \\ \hline\hline
 \end{tabular}
\label{Tab3}
\end{table}
\end{center}

\subsection{Bounds from $\beta$ to $\bar{s}^{jk}$}

It is however clear that relation (\ref{betaftrace}) can be formally inverted, resulting in $\delta \simeq 3 (\mpl/2\pi M)^2 \beta$, and suggesting therefore also the opposite path, namely the possibility to transfer bounds from  $\beta $ to $\bar{s}^{jk}$. Obviously, in this context we keep considering only the bounds on $\beta$ displayed in Table \ref{Tab2}, i.e. those of gravitational origin.

$\bullet$ If we consider the bound on $\beta$ coming from the perihelion precession in the solar system, $\beta < 10^{69}$, then we use  $M=M_{\bigodot}\simeq 10^{38}\mpl$, and from the above relation we get $|\bar{s}^{jk}|<10^{-9}$, perfectly in line with the correspondent value displayed in Table \ref{Tab01}.

$\bullet$ If we consider the bound on $\beta$ coming from the gravitational wave event GW150914, $\beta < 10^{60}$, then we should use for $M$ the total mass of the (supposed) two-black holes system, roughly $M=m_1+m_2\simeq 50 \,M_{\bigodot}\simeq 5\times10^{39}\mpl$. From the above relation we the obtain $|\bar{s}^{jk}| \lesssim 10^{-21}$. Obviously, this evaluation cannot be retained as reliable as the previous one, given for example the still large uncertainties affecting the event GW150914.

$\bullet$ The constraint $\beta < 10^{21}$ coming from universality of free fall, or from the law of reciprocal action, is somehow problematic in this context. Such constraint is derived in Ref.\cite{ghosh} by postulating a deformation of the classical (covariant) Poisson brackets which resembles the deformed quantum commutator (\ref{gup}). As it is explicitly shown in Ref.\cite{SC2}, a deformation of Poisson brackets implies immediately a deformation of the equation of motion (i.e. of the geodesic equation, in the relativistic context), in such a way that the motion of a test particle depends on the mass of the particle itself. That is, a violation of the equivalence principle. This immediately reflects on the modified relations for free fall, or for reciprocal actions, from which in Ref.\cite{ghosh} the bound $\beta < 10^{21}$ is obtained. On the contrary, the other bounds reported in Table \ref{Tab2} are derived by deforming the dispersion relation (Ref.\cite{feng17}), or by deforming the metric (Ref.\cite{SC2}), in order to mimic the GUP-deformed Hawking temperature, but this is done always under the strict assumption of validity of equivalence principle, i.e. without deforming the geodesic equation.

 Because of its origin from violation of the equivalence principle, the bound $\beta<10^{21}$ should be therefore considered not homogeneous with the others reported in Table \ref{Tab2}, which on the contrary respect the equivalence principle. Moreover, the bound $\beta < 10^{21}$, with the use of $M=M_{Earth}$, would imply $|\bar{s}^{jk}| < 1.01 \times 10^{-45}$. This suppresses by many orders of magnitude the current bounds reported in Table \ref{Tab01}, and appears quite unexpected with respect to the corresponding bounds of SME coefficients in the matter sector, where the present sensitivity of experiments has not provided the evidence of such a suppression (in the SME, in fact, coefficients can be transferred from the gravitational sector to matter sector and viceversa with an appropriate change of coordinates).
\footnote{The bound $\beta < 10^{21}$ can perhaps be made more reasonable if we consider that, according to \cite{Tkachuk}, where composite systems have been investigated, to get the deformation parameter referred to particle physics one should consider an effective deformation parameter that must be multiplied with the square of the nucleon number $N_{nuc}$ making up the gravitational source.
Considering the Moon with $M_{Moon}\simeq 7.3\times 10^{22}$Kg and the nucleon mass $m_{nuc}\simeq 1.67\times 10^{-27}$Kg, one gets $N_{nuc}=\frac{M_{Moon}}{m_{nuc}}\simeq 4\times 10^{49}$. The deformation parameter turns out to be $\beta=N_{nuc}^2 10^{21}$, and therefore this procedure increases the deformation parameter by many orders of magnitude, relaxing in such a way the strong suppression of the SME parameter $s^{jk}$. It should also be noticed, however, that the considerations  proposed in Ref.\cite{Tkachuk} are themselves based on deformed Poisson brackets, namely on violation of equivalence principle.}

In this context, it is however instructive to derive an upper bound on $\bar{s}^{jk}$ by using directly the present bounds on the violation of Law of Reciprocal Action obtained from data of Lunar Laser Ranging. According to \cite{bondi,lamm} the active mass $m_a$ is the source of the gravitational field ($\nabla^2 V({\bf x})=-4\pi m_a \delta{{\bf x}}$), while the passive mass is related to the response of a mass to a gravitational field, and appears in the equation of motion $m_i {\ddot {\bf x}}=m_p \nabla V({\bf x})$, where $m_i$ is the inertial mass. Following \cite{ghosh,lamm}, the equations of motion for a gravitationally bound system of particles, and of its center of mass ${\bf X}$ are
 \[
 m_{1i} {\ddot {\bf x}_1}=Gm_{1p} m_{2a}\frac{{\bf x}_2-{\bf x}_1}{|{\bf x}_2-{\bf x}_1|^3}\,,
 \]
 \[
 m_{2i} {\ddot {\bf x}_2}=Gm_{2p}m_{1a}\frac{{\bf x}_1-{\bf x}_2}{|{\bf x}_1-{\bf x}_2|^3}\,,
 \]
 \[
 {\ddot {\bf X}}=G\frac{m_{1p} m_{2p}}{m_{1i}+m_{2i}} C_{21}\frac{{\bf x}}{|{\bf x}|^3}\,, \quad C_{21}=\frac{m_{2a}}{m_{2p}}-\frac{m_{1a}}{m_{1p}}\,,
 \]
where ${\bf x}$ is the relative coordinate. If $C_{21}\neq 0$, then the center of mass possesses a self-acceleration. In SME model this occurs through the coefficients $\bar{s}_n^{jk}$ by assuming that they are particle depending. Here, therefore, $m_{ni}$, $(n=1, 2)$ is the inertial mass of particles, and $m_{na}=(1+\bar{s}_n^{jk}{\bf {\hat x}^j}{\bf {\hat x}^k})m_{ni}$ is the active mass of particles (see Eq.\ref{LV2}).  Besides, $m_{np}$ is identified with $m_{ni}$. The absence of self interaction of the Moon (one considers the distribution of the main constituent of Moon, $Al$ and $Fe$) yields the bound $|C_{Al-Fe}|<7\times 10^{-13}$, which implies $|\delta_{Al-Fe}|< 7\times 10^{-13}$, which is in clear agreement with the expected bounds for $\bar{s}^{jk}$ of Table \ref{Tab01}.

$\bullet$ In Ref. \cite{SLV} the deformation parameter $\beta$ has been computed  by making use of the leading quantum corrections to the Newtonian potential \cite{Duff,Dono,Dono2}. The corrections to the Newtonian potential imply naturally a quantum correction to the Schwarzschild metric, and this leads to a precise numerical value for $\beta$, namely $\beta = 82\pi/5$, which is of the same order of magnitude expected from string theory. If we use this value for $\beta$, together with the Earth mass, $M\simeq 2.74 \times 10^{32}\mpl$, we get $|\bar{s}^{jk}| \simeq 5.21 \times 10^{-65}$, a value which is hugely beyond the tested experimental bounds of $\bar{s}^{jk}$. Although the actual physical relevance of this bound remains questionable, such a minute value is somehow expected, since with the above procedure we checked, for the first time, the Lorentz violating SME coefficients against a quantity, $\beta$, typically linked with Planck scale phenomena. The previous gravitational bounds on ${\bar s}^{jk}$ are all obtained in the contest of classical gravity, although post Newtonian. Here instead we face a "quantum" gravity scenario, or to be more precise, a semiclassical gravity scenario, represented by the Hawking effect. To this circumstance can be presumably traced back such a huge refinement of the value of the $\bar{s}^{jk}$ coefficients.

\section{Conclusions}

In this paper we have derived an upper bound on the deformation parameter $\beta$ of the generalized uncertainty principle (\ref{gup}), by relating $\beta$ to the coefficients ${\bar s}^{jk}$ defined in the gravitational sector of SME.
The main point of the derivation relies on the fact that we directly compute a quantum mechanical effect, the Hawking temperature, for which the GUP is necessarily relevant, without postulating a specific representation of canonical commutators. We then compute the same temperature using a deformed Schwarzschild-like metric, thereby linking together the deformed uncertainty relation, and the deformed metric. It is noteworthy that in our formalism General Relativity and standard Quantum Mechanics are recovered in the limits $s^{jk}\to 0$ and $\beta\to 0$, respectively.

Our main results can be summarized in two distinct cases:

\begin{itemize}
  \item By considering the experimental upper bounds on the parameter $|{\bar s}^{jk}|<10^{-15}$, we infer a bound on the GUP deformation parameter $\beta<10^{51}$, which improves by many order of magnitude the bounds obtained in gravitational frameworks compatible with the equivalence principle, and it lies quite close to the Landau level measurements, obtained with non-gravitational measurements/experiments.
  \item If we adopt for the parameter $\beta$ the value $82\pi/5$, derived in the framework of QFT and GR, then the coefficients ${\bar s}^{jk}$ turns out to be bounded by ${\bar s}^{jk}\simeq 10^{-65}$. Such a minute value, although extremely tight because derived within a semiclassical gravity approach (Hawking effect), seems however to demand for further investigations, both on the experimental as well on the theoretical side.
\end{itemize}

There is, nowadays, a lively debate on the measurable features implied by various kinds of GUPs, and many efforts are devoted to the predictions about the size of these modifications.
In this respect, several experiments have been also proposed to test GUPs in the laboratory.
As shown in this paper, GUP measurements could have an interplay with the violation of the fundamental symmetries in physics, such as CPT and Lorentz invariance, through the SME. Here we focus on SME for the gravitational sector, but to understand whenever the other coefficients of the model may affect GUP, or specific representations of canonical operators, might be of great interest, especially in perspective of possible links with quantum gravity.
These aspects, in turn, are particularly appealing in view of the possibility to create, in the next future, a laboratory-scale imitation of a black hole horizon, emitting analogue Hawking radiation \cite{ron}.

\section*{Acknowledgements} The authors thank A. Kostelecky for helpful suggestions and enlightening comments.

\appendix
\section{Solution of equation (\ref{eqr})}
To solve Eq.(\ref{eqr}), first we formulate it in a general form
\be
x = a + \ve f(x)~.
\label{eqx}
\ee
It is obvious that if $\ve$ is set equal to zero, then the solution will be $x_0=a$.  If $\ve$  is slightly different from zero,
then we can try a test solution of the form $x_0 = a + \eta(\ve)$
where $\eta(\ve) \to 0$ for $\ve \to 0$. Substituting the aforesaid test solution in (\ref{eqx}), we get $x_0 = a + \ve f(x_0)$
which means $\eta = \ve f(a+\eta)$.
To first order in $\eta$, we have $\eta = \ve[f(a) + f'(a)\eta]$ from which we obtain
$\eta = \ve f(a)/[1-\ve f'(a)]$.
Therefore, to first order in $\ve$, the general solution of (\ref{eqx})  reads
\be
x_0 \ = \ a + \frac{\ve f(a)}{1-\ve f'(a)}.
\ee
%

\section{Expansion in $\ve$ of $T$}
Hawking temperature is given by
\be
T =  \frac{\hbar}{4\pi}F'(r_H)~.
\label{hawtemp}
\ee
From Eq. (\ref{DM}), one gets
\be
F'(r) = \frac{a}{r^2} \ + \ \ve\,\phi'(r)\,.
\ee
It is useful to write the solution (\ref{solr}) in the compact form $r_H = a(1-\lambda)$
where $\lambda = \ve\, \phi(a)/\{1 \ + \ \ve\,[\phi(a) \ + \ a\,\phi'(a)]\}$  and,
therefore, $\lambda \sim \ve$, $|\lambda| \ll 1$.
Then
\be
F'(r_H) = \frac{1}{a(1-\lambda)^2} \ + \ \ve \phi'[a(1-\lambda)]\,.
\ee
Expanding in $\ve$ this last expression, one gets Eq.(\ref{teps}).

%
%
%
\end{document}